# Fully Automated Verification Framework for Configurable IPs: From Requirements to Results

Shuhang Zhang, Infineon Technologies AG, Munich, Germany (*Shuhang.Zhang@infineon.com*)

Jelena Radulovic, Infineon Technologies d.o.o. Belgrade, Serbia (*Jelena.Radulovic@infineon.com*)

Thorsten Dworzak, Infineon Technologies AG, Munich, Germany (*Thorsten.Dworzak@infineon.com*)

*Abstract*—The increasing competition in the semiconductor industry has created significant pressure to reduce chip prices while maintaining quality and reliability. Functional verification, particularly for configurable IPs, is a major contributor to development costs due to its complexity and resource-intensive nature. To address this, we propose a fully automated framework for requirements-driven functional verification. The framework automates key processes, including vPlan generation, testbench creation, regression execution, and reporting in a requirements management tool (RMT), drastically reducing verification effort. This approach accelerates development cycles, minimizes human error, and enhances coverage, offering a scalable and efficient solution to the challenges of verifying configurable IPs in today's competitive market.

I. INTRODUCTION

The semiconductor industry has become increasingly competitive in recent years, driven by rapid advancements in technology and a growing demand for innovative, high-performance, and cost-effective solutions. This competition has created immense market pressure to reduce the price of manufactured chips while maintaining, or even improving, their quality and reliability. With the proliferation of applications like artificial intelligence, automotive systems, and Internet-of-Things (IoTs) devices, there is a constant push for faster time-to-market and reduced development costs. As a result, semiconductor companies are striving to optimize every stage of the design and manufacturing process to meet these challenges.

One of the most resource-intensive phases of chip development is functional verification. Verification accounts for a significant portion of the overall design cycle, often consuming up to 70% of the total development effort. This is especially true for configurable Intellectual Properties (IPs), which add another layer of complexity to the process. Configurable IPs allow designers to tailor functionality to specific use cases, offering flexibility and reusability. However, this configurability translates to a vastly increased number of design permutations, each of which requires rigorous verification to ensure compliance with specifications and reliable functionality. The verification required for these IPs inflates both development time and cost, which ultimately contributes to higher prices for the final chip.

To address these challenges, automation in functional verification has emerged as a promising solution. By automating repetitive and labor-intensive tasks, verification teams can focus their efforts on higher-value activities, such as debugging and architectural validation. Automation not only accelerates the verification process but also ensures consistency, reduces human errors, and enhances coverage. As semiconductor designs grow in complexity, the need for more sophisticated and automated verification solutions has become increasingly apparent. This trend aligns with broader industry goals of reducing costs and accelerating development cycles while maintaining product quality.

In this work, we propose a fully automated framework for requirements-driven functional verification of configurable IPs. The framework is designed to streamline the verification process by automating key stages, including the generation of a verification plan (vPlan), the creation of testbenches, regression execution, and RMT reporting. By eliminating much of the manual effort typically associated with these tasks, the framework achieves significant reductions in both verification time and resource usage. In this work, we demonstrate the proposed framework using a configurable memory sub-system IP as an example. This IP supports multiple safety mechanisms, various low power modes and a range of memory technologies. By employing our framework, the verification effort for each configuration is reduced significantly, from 40 person-days (PD) to just 2 PD.

II. BACKGROUND

*A. Configurable IPs*

Configurable IPs in digital design play a crucial role in building flexible and reusable hardware components, allowing designers to tailor functionality to specific application requirements efficiently. The general workflow of a configurable IP can be described as follows: given a specific configuration, the corresponding Register Transfer Level (RTL) can be generated using the configurable IP, as illustrated in Figure 1. Configurable IPs can be implemented through two primary methods: parameterized SystemVerilog modules and dedicated IP generators. Parameterized SystemVerilog modules enable flexibility by allowing users to define design parameters, such as data widths and the base address, at compiling time. This approach avoids the need to manually rewrite or redesign modules for different use cases, promoting code reuse and reducing development overhead. Additionally, with parameterization, a single module can be extensively verified, ensuring reliability across multiple configurations. On the other hand, IP generators provide a more automated and user-friendly approach to configurable designs. These tools allow designers to generate IP blocks through graphical user interfaces or scripts. They offer advanced features, including automated clock domain handling, interface generation, and performance optimization, significantly accelerating time-to-market while ensuring design consistency. Together, these approaches make configurable IPs indispensable for modern, scalable digital design workflows.

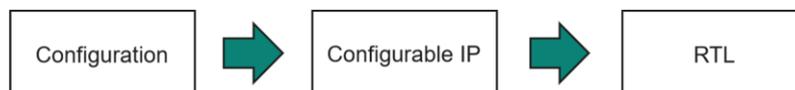

Figure 1. General workflow of a configurable IP.

*B. Related Works*

Automation in functional verification has emerged as a transformative solution to address the growing complexity and challenges associated with verifying configurable IPs. Configurable IPs, by nature, demand extensive verification to ensure functionality across a wide range of configurations, which can become a resource-intensive and time-consuming process. To mitigate this, recent advancements in automation have primarily concentrated on the execution phase of verification, yielding substantial progress. For instance, the adoption of configurable and reusable testbenches has become a standard practice, enabling efficient validation across multiple configurations while minimizing redundant effort [1-2]. These testbenches can be parameterized and reused for different configurations of the same IP, enhancing productivity and consistency. Similarly, automated formal verification frameworks have been leveraged to address specific challenges posed by highly configurable IPs. Formal methods use mathematical proof techniques to verify properties against all possible configurations, significantly reducing manual effort and improving confidence in IP correctness [3-6].

Despite this progress, there remains a lack of a fully automated, end-to-end verification flow that seamlessly integrates all phases of the verification lifecycle. A robust automated flow would establish a closed loop from requirements capture, to testcase generation, to verification, and back to requirement validation, ensuring a comprehensive and traceable process. Bridging this gap is essential for further advancing automation in functional verification.

III. PROPOSED VERIFICATION FRAMEWORK

For configurable IPs, we have developed a fully automated end-to-end flow that encompasses all key stages, including concept, design, and verification, as illustrated in Figure 2. The customer, or project team, can configure and order the IP through a web interface, selecting from various options such as memory technologies. These selected options are then exported into a specification Excel file, which serves as the basis for downstream processes. The design team uses this file to generate RTL code, while the concept team utilizes it to create hardware requirements (HWRQs) within the RMT database (We use Jama as an example). As this work focuses on the verification process, the implementation details related to the design and concept stages are not covered in depth.

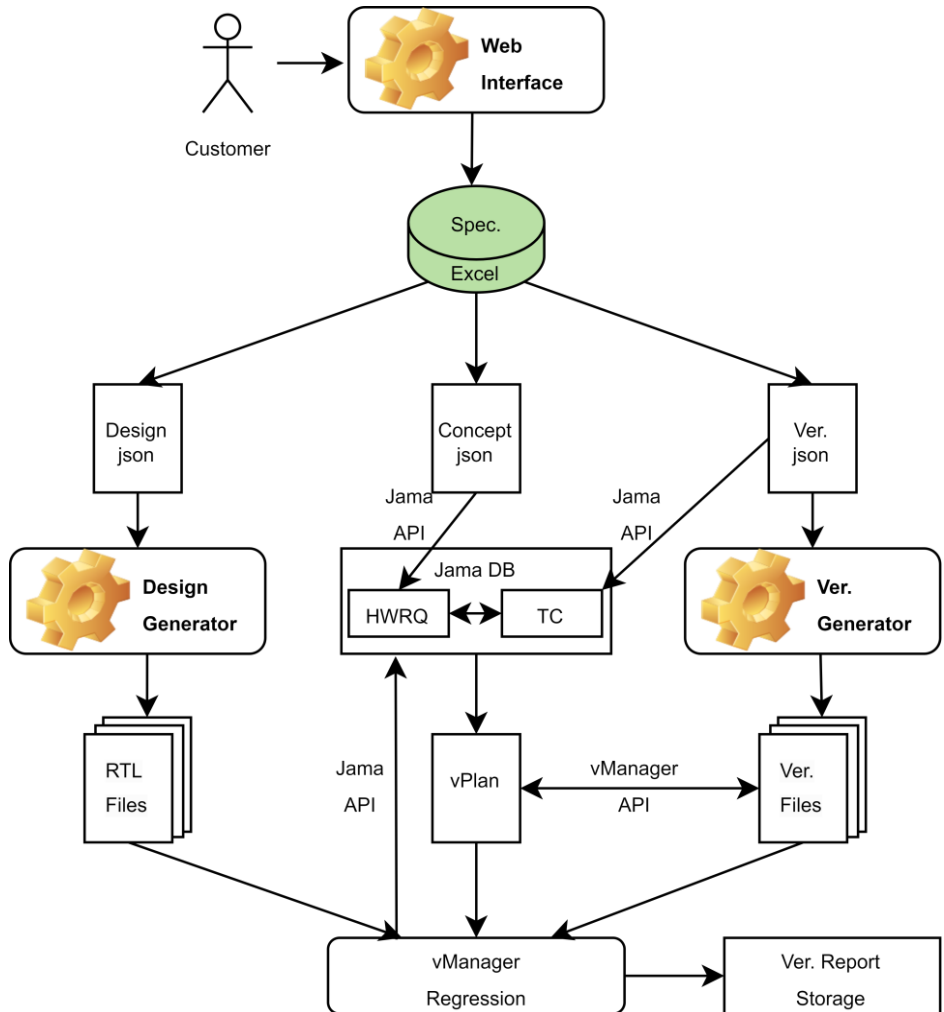

Figure 2. Fully Automated Development Framework for Configurable IPs: Once the customer places an order for the IP, all subsequent steps are triggered automatically. Upon completion of the process, the customer receives the IP design files and verification results, delivered seamlessly and automatically.

To address the challenges of functional verification for configurable IPs, we propose a fully automated verification framework. This framework significantly minimizes manual effort by automating critical stages of the verification process, from requirement management to regression execution and reporting. The framework is divided into four steps: Jama DB generation, vPlan generation, testbench generation and regression, and automated reporting to Jama. Each step leverages different APIs and automation techniques to simplify and streamline the verification workflow. Below, we describe each step in detail.

A. *Jama Database Generation*

The first step in the framework focuses on automating the creation of the Jama database, which serves as the central repository for managing hardware requirements and corresponding verification testcases. Based on customer inputs—such as specifications, functional requirements, and configuration details—the framework generates the necessary content in the Jama database automatically. This includes the creation of testcases aligned

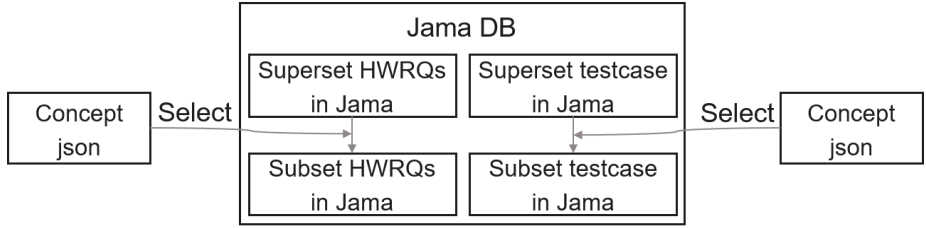
Figure 3. Creation of the Jama database using the Jama API.

with the specified hardware requirements. Relationships between hardware requirements, testcases, and verification domains are also established programmatically using the Jama Application Programming Interface (API). By automating this process, we eliminate the need for manual data entry, ensuring consistency and traceability while significantly reducing human errors. This ensures that the initial groundwork for verification is both accurate and comprehensive.

Figure 3 illustrates the creation of the Jama database using the proposed framework. In this work, we begin by drafting supersets of Hardware Requirements (HWRQs) and Testcases within Jama, ensuring they comprehensively cover all possible features and configurations of the configurable IP. These supersets are intended to act as a complete repository of requirements and testcases, accounting for every potential variation. To ensure quality and correctness, the supersets, along with the relationships between HWRQs and Testcases, undergo an intensive review process. Once the specific configurations for the IP are defined by the project team, corresponding subsets of HWRQs and Testcases are generated from these supersets. These subsets are tailored to the defined configurations and represent only the relevant requirements and tests necessary for that specific instance. Subsets are created programmatically using the Jama API, which facilitates automated selection of items from the supersets and their generation as new items in Jama. This process can be efficiently implemented using the post_item() function from the Jama API, streamlining the subset creation process and ensuring consistency across different configurations.

B. *vPlan Generation*

The second step in the process involves the automated generation of the verification plan (vPlan), which functions as a structured roadmap for the functional verification of the configurable IP. The vPlan captures critical elements such as verification goals, coverage criteria, and associated testcases, ensuring strong alignment between hardware requirements and the verification process. Starting with the hardware requirements specified by concept engineers and the test cases created in the earlier step, the framework first generates a detailed XML file, referred to as ipvs.xml in Figure 4, which consolidates all relevant information stored in the Jama database. This XML file serves as the foundation for the vPlan generation. Using an in-house vPlan generator (traceability), the framework processes the ipvs.xml file to automatically generate a complete vPlan file. At this stage, no explicit mapping exists between the vPlan items and the SystemVerilog verification items.

To bridge this gap, the framework further integrates the vPlan with SystemVerilog items through the vManager API. This is achieved by employing the add_mapping_pattern() function, which establishes associations between vPlan items and corresponding SystemVerilog verification entities, such as assertions, functional coverage points, and test cases. This mapping step is critical for ensuring traceability between high-level requirements and low-level verification implementation. By automating both the vPlan creation and its mapping to SV, the framework guarantees that verification goals are comprehensive, measurable, and directly tied to the requirements. This approach enhances transparency, improves traceability, and reduces manual effort, streamlining the verification process and ensuring consistency across all stages.

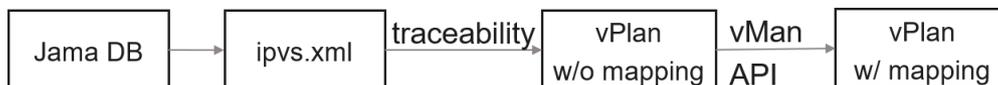
Figure 4. Creation of vPlan using an in-house tool and the vManager API.

*C. Testbench Generation and Regression*

In the third step, the framework focuses on automating the generation of testbenches and regression execution tailored to the specific IP configuration being verified. This comprehensive automation ensures that the verification environment—including both simulation-based and formal verification setups—is customized to the unique functional requirements and characteristics of the IP. The goal of this step is to streamline the preparation and execution of verification, significantly reducing manual efforts while ensuring consistency and reliability.

*1)* Simulation-Based Verification

For simulation-based verification, the framework leverages an in-house testbench generator to create the skeletal Unified Verification Methodology (UVM) testbench. However, the output of the in-house generator is incomplete and cannot be directly used for verification execution. The framework addresses this limitation by generating the missing components required to complete the testbench. Specifically, it automates the creation of critical UVM components such as testcase sequences, scoreboards, drivers, and monitors, ensuring these elements are consistent with the IP's functional requirements. Additionally, other essential files for simulation, such as property files and Unified Power Format (UPF) wrapper files, are also generated to support more comprehensive validation of the IP. Alongside these elements, the framework prepares the simulation setup by generating simulation options files and Makefiles, which are necessary to configure and run the simulation environment efficiently. These setup files ensure that the simulation flow is seamless and aligned with specific IP configurations, enabling effective and repeatable execution of test cases.

*2)* Formal Verification

For formal verification, the framework simplifies the setup process even further, given the deterministic nature of formal methods. The primary focus here is on generating property files that define the assertions and properties to be formally verified. These property files are automatically created based on the project-defined IP configuration, ensuring that the verification process targets the correct functional requirements. Since formal verification environments are less complex than simulation-based setups, this automation accelerates the preparation process, allowing verification engineers to focus on analyzing results and debugging.

*3)* Regression Setup

In addition to automating the creation of simulation and formal verification environments, the framework also handles the preparation for regression execution. Both simulation- and formal-based regression workflows require specific configuration files and scripts to enable automated execution within the vManager environment. To facilitate this, the framework generates verification setup input files (VSIF) and Tcl scripts, which are essential for scheduling, managing, and tracking regressions in vManager. These files ensure that all testcases—whether simulation or formal—are executed systematically across the defined configurations of the IP.

*4)* Regression Execution

Once the testbench or formal verification environment is ready, the framework automatically triggers regression execution. It systematically runs all testcases and properties defined for the specific IP configuration. If any failures or discrepancies occur during regression, the framework immediately notifies the verification team by providing detailed failure logs and reports, enabling prompt and efficient debugging.

By automating the generation of testbenches, simulation-based or formal verification setups, and regression configurations, this framework significantly reduces manual effort, improves traceability, and ensures that verification environments are tailored to the specific needs of configurable IPs. This end-to-end automation streamlines the entire verification process, increases productivity, and enhances the overall quality of the IP under development.

*D. Reporting to Jama*

The final step in the framework is dedicated to reporting verification results back to the Jama database, ensuring complete traceability and transparency across the verification process, which is shown in Figure 5. Once

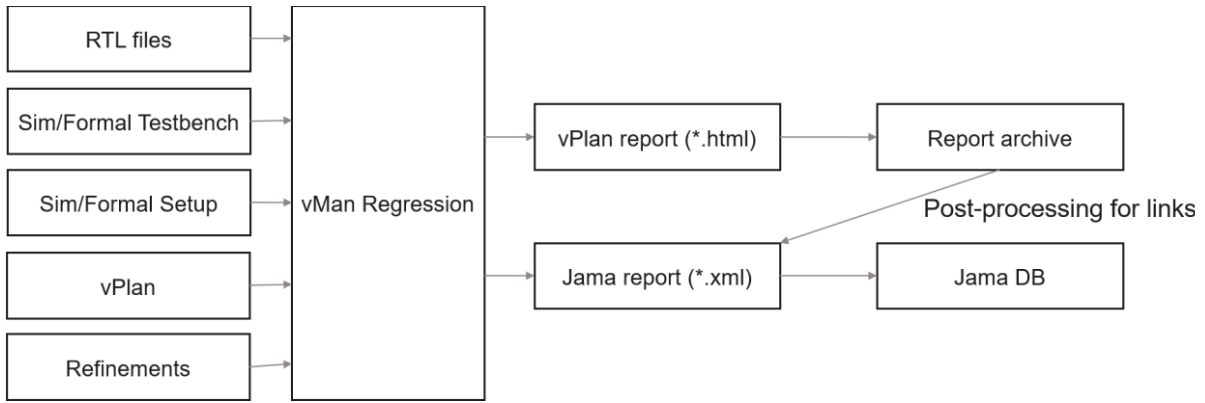
Figure 5. Extract regression results and push results to Jama.

the regression phase is complete, the framework seamlessly collects and organizes the verification results using Tcl scripts and the vManager API. These results include critical data such as test pass/fail statuses and detailed coverage metrics, making it easier for stakeholders to assess the overall quality and completeness of the verification process. To achieve this, the framework generates two types of reports: the HTML vPlan report and the XML Jama report, each serving distinct purposes.

The HTML vPlan report is produced by vManager, a widely adopted tool, and provides a detailed summary of the coverage achieved for all pre-defined items in the verification plan. This report is instrumental in understanding how well the defined verification goals—such as functional coverage points, assertions, and testcase execution—have been met. The second report, the XML Jama report, is specifically generated by the framework to update the testcase items inside Jama. This report captures the pass/fail statuses and coverage results in a structured format that aligns with the requirements and test cases stored in Jama. Using the Jama API, the XML Jama report is pushed back into the database, where the results are linked directly to the corresponding requirements and test cases, ensuring end-to-end traceability.

To further enhance user accessibility and review efficiency, the framework addresses a common pain point: the difficulty of accessing the HTML vPlan report from within the server workspace. To make it more reviewer-friendly, the framework uploads the HTML vPlan report to an internal archive where it is securely stored. A link to this archived report is then added to the XML Jama report. When the XML Jama report is pushed back into Jama, this link becomes readily available within the verification report section of each testcase item. This approach ensures that reviewers can conveniently view the comprehensive HTML vPlan report directly from Jama, eliminating the need to navigate server directories or external tools.

By automating the generation, organization, and reporting of verification results back into Jama, the framework significantly streamlines the review process and ensures all team members have easy access to the verification outcomes. This step not only guarantees complete traceability between requirements, testcases, and verification results but also fosters transparency and collaboration across teams. Ultimately, this robust reporting mechanism allows stakeholders to quickly assess whether all requirements have been successfully verified, further enhancing the efficiency and reliability of the overall verification process.

IV. FRAMEWORK ROBUSTNESS

The robustness of the proposed framework is a cornerstone, ensuring reliable and consistent performance across all stages of the fully automated verification flow. In any highly automated process, robustness is critical to minimizing errors, maintaining traceability, and achieving high-quality verification outcomes. The framework achieves this robustness through a combination of rigorous methodologies and thorough validation steps, shown in Figure 6. One key factor contributing to its reliability is the extensive testing of configurations, which enhances the framework's ability to handle a wide range of scenarios. By integrating automated flows tailored to configurable IPs, the framework is designed to accommodate the inherent complexities of these designs, ensuring that it can seamlessly generate and execute verification plans, environments, and regressions for diverse configurations. The ability to dynamically adapt to varying design requirements is a fundamental strength of the framework, further solidifying its robustness.

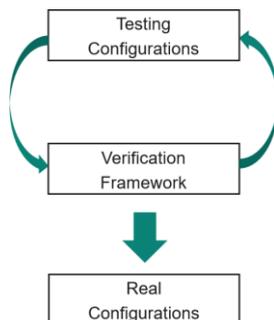

Figure 6. Robustness of the framework ensured through validation configurations and comprehensive reviews.

Another significant contributor to its robustness is the comprehensive review process applied during the creation of superset Hardware Requirements (HWRQs) and TestCases (TCs). These supersets undergo an intensive review to ensure they are complete, accurate, and effectively cover all potential features of the configurable IP. The relationships between HWRQs and TCs are also meticulously reviewed, ensuring that the traceability between requirements and test cases is unambiguous and reliable. This thorough preparation of supersets forms the foundation for generating tailored subsets, which are critical for successful verification of specific configurations.

Additionally, the framework incorporates detailed waiver reviews as part of its robustness strategy. Waivers play an essential role in managing exceptions or deviations in the verification process, and the framework's rigorous review of these waivers ensures that potential issues are caught early and addressed appropriately. This process minimizes the risk of oversight and guarantees that any deviations from the standard flow are well-documented and justified.

By combining extensive testing configurations, meticulous reviews of superset requirements and test cases, and a structured approach to waiver evaluation, the proposed framework establishes a robust foundation for fully automated verification flows. This robustness ensures that the framework can deliver reliable verification results across a wide spectrum of IP configurations, fostering confidence in its ability to handle the complexities of modern digital design. Ultimately, the framework's robustness not only enhances its usability and scalability but also contributes to reducing debugging efforts, improving overall productivity, and ensuring the delivery of high-quality IPs.

V. RESULTS

In this work, we demonstrate the proposed framework using a configurable memory subsystem IP as an example, shown in Figure 7. This memory subsystem is divided into two main components: the logic memory part and the physical memory part. The logic memory part includes an AHB interface, which decodes AHB signals into internal access signals to perform memory read and write operations. Additionally, this block houses a configurable Error Detection and Correction (EDC/ECC) module, capable of supporting functionalities ranging from single error detection to double error correction and triple error detection, offering robust error management for the memory subsystem. On the other hand, the physical memory part primarily comprises memory arrays and supports various memory technologies and low-power modes. The high configurability of this IP, with multiple parameters and functional features, introduces significant challenges for its verification.

To address these challenges, we adopt a hierarchical verification approach, utilizing both simulation and formal verification techniques tailored to the unique characteristics of each component. The logic memory wrapper, being a fully digital block, is verified entirely using formal verification. Formal techniques are particularly suitable for this block due to its deterministic nature and the inherent complexity of the configurable EDC/ECC module, allowing exhaustive validation of all functional scenarios. For the physical memory wrapper, which involves analog and mixed-signal behaviors, a simulation-based testbench is employed. At this level, the testbench directly drives internal access signals to validate read and write operations effectively.

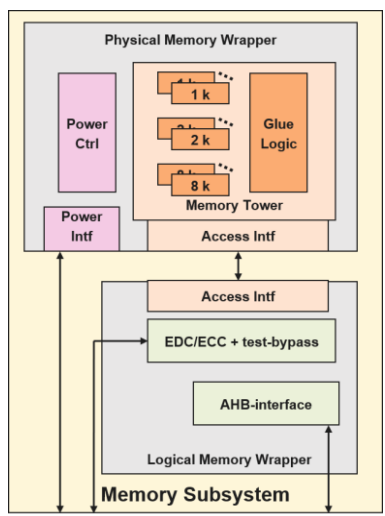

Figure 7. A configurable memory subsystem.

At the subsystem level, we take a broader approach by utilizing a simulation-based testbench integrated with an AHB Verification IP (VIP). This enables testing of the complete memory subsystem, including the interaction between the logic and physical memory parts, as well as verifying support for all AHB protocol features and low-power modes. The hierarchical verification methodology allows us to identify bugs early at the sub-block level, ensuring that each component functions correctly before integration. The subsystem-level testbench then provides additional confidence by verifying the overall functional correctness and integration of the memory IP.

We applied the proposed framework to generate both simulation and formal verification environments for this configurable memory subsystem IP. Using this flow, we successfully verified over 100 different configurations within a short period, identifying several bugs in the process. Most of these bugs were related to corner cases in specific configurations, which are particularly challenging to detect using traditional approaches. By leveraging the automation provided by the framework, the verification effort for each configuration was significantly reduced—from an estimated 40 person-days (PD) down to just 2 PD. This demonstrates the framework's ability to streamline the verification process, improve efficiency, and ensure thorough coverage even for highly configurable designs.

## VI. CONCLUSION

Based on our project, the automated verification framework demonstrates a substantial reduction in functional verification effort by streamlining the generation, execution, and reporting processes. By automating tasks such as testbench creation, regression setup, and result reporting, the framework minimizes manual intervention, reduces errors, and enhances overall productivity. Its ability to ensure traceability from requirements to verification results further improves verification quality and reliability. Given its success in this project, this workflow is planned to be extended to additional projects in the future.

## REFERENCE

[1] G. Auditore and G. Falconeri. Automating the formal verification sign-off flow of configurable digital IP's. DVCon, 2019.
[2] R. Misra, S. Rao, A. Kumar, G. Srivastava, Y. Kim, S. Choi. Building Confidence in System level CPU Cache Coherency Verification for Complex SoC's through a Configurable, Flexible and Portable Testbench. DVCon, 2022.
[3] K. Krishna and S. Veerapur. Configurable Testbench (TB) for Configurable Design IP. DVCon, 2022.
[4] S. Zhang, B. Olmos and B. Naik. Automated Formal Verification of a Highly-Configurable Register Generator. DVCon, 2024.
[5] S. Zhang and B. Olmos. Automated Formal Verification of Area-Optimized Safety Registers in Automotive SoCs. DVCon, 2025.
[6] S. Zhang, B. Olmos, W. Kunz, and D. Lettnin. Synthesis-Aware Area Optimization for Safety Registers in Automotive SoC. IEEE Computer Society Annual Symposium on VLSI (ISVLSI), 2025.